\documentclass[10pt,aps,prl,twocolumn,superscriptaddress,showpacs]{revtex4-1}

\usepackage[pdftex]{graphicx}
\usepackage{bbm}
\usepackage{bbold}
\usepackage{amssymb}
\usepackage{amsmath}
\usepackage{xspace}
\usepackage{color}
\usepackage{mathtools}

\begin{document} 

\title{Anyonic Haldane insulator in one dimension}

\author{Florian Lange}
\affiliation{
Computational Condensed Matter Physics Laboratory, RIKEN, Wako, Saitama 351-0198, Japan}
\affiliation{Institut f\"ur Physik, Ernst-Moritz-Arndt-Universit\"at
Greifswald, 17489 Greifswald, Germany}
\author{Satoshi Ejima}
\affiliation{Institut f\"ur Physik, Ernst-Moritz-Arndt-Universit\"at
Greifswald, 17489 Greifswald, Germany}
\author{Holger Fehske}
\affiliation{Institut f\"ur Physik, Ernst-Moritz-Arndt-Universit\"at
Greifswald, 17489 Greifswald, Germany}

\date{\today}

\begin{abstract}
We demonstrate numerically the existence of a nontrivial topological
 Haldane phase for the one-dimensional extended ($U$-$V$) Hubbard model
 with a mean density of one particle per site, not only for bosons but also for anyons, 
 despite a broken reflection parity symmetry. The Haldane insulator, surrounded by superfluid, Mott insulator and density-wave phases in the  $V$-$U$ parameter plane,  is protected by combined (modified) spatial-inversion and time-reversal symmetries, which is verified within our matrix-product-state based infinite density-matrix renormalization group scheme by analyzing generalized transfer matrices.   With regard to an experimental verification of the anyonic Haldane insulator  state the calculated asymmetry of the dynamical density structure factor should be of particular importance.
\end{abstract}

\maketitle

Anyons represent a third fundamental class of particles with fractional exchange 
statistics that interpolates, to some degree, between those of bosons and fermions 
having symmetric or antisymmetric wave functions under exchange~\cite{LM77,Wi82}. By contrast, the
exchange of two anyons creates a phase factor $e^{\mathrm{i}\theta}$ in the many-body 
wave function, where the statistical parameter $\theta$ can be of any value in the interval $(0,\pi)$.   
In the beginning anyons were thought to be relevant solely for two-dimensional  
systems. Describing the fractional quantum Hall effect experiments in particular, the
quasiparticles could be viewed as anyons with $\theta$ fixed  by the filling factor~\cite{TSG82,Laughlin83}. 
With Haldane's generalized Pauli principle and definition of fractional statistics, however, the concept of anyons 
becomes important in arbitrary dimensions~\cite{Haldane91}.

In one dimension, the physics of anyons might be studied successfully with ultra-cold atoms 
in optical lattices ~\cite{BDZ08}.  For example, 
one-dimensional (1D)  anyon statistics can be implemented by bosons with occupation-dependent hopping amplitudes generated 
by assisted Raman tunneling~\cite{KLMR11,GS15}.
An alternative route to create 1D anyons in an optical lattice 
exploits lattice-shaking-assisted tunneling against 
potential offsets generated by a combination of a static potential tilt
and strong on-site interactions~\cite{SSE16}. Thereby, advantageously, no additional 
lasers are required, except for those employed on creating optical lattices.
However, in spite of the huge experimental efforts, a conclusive detection  
of 1D anyons in optical lattices has not yet been achieved.

Notwithstanding, from a theoretical point of view, anyons in one dimension
have received continuous and legitimate interest on account
of their intriguing physical properties. 
The exact solution  of an 1D anyon gas with delta-function potential
has been obtained by  a Bethe ansatz technique~\cite{Kundu99}.
Boundary conformal field theory shows  that  non-Abelian anyons
may form topological insulating phases in spin-1/2 su(2)$_k$ chains~\cite{DeGottardi14}.
For the Abelian 1D anyon-Hubbard model (AHM), 
the possibility of a statistically induced quantum phase transition
between Mott-insulator (MI) and superfluid phases~\cite{KLMR11,AFS16}
and the asymmetry of the momentum distribution for hard-core~\cite{HZC09}
and soft-core anyons~\cite{TEP15} have been addressed so far.  
Since the AHM is equivalent to a variant of the Bose-Hubbard model (BHM)
with state-dependent bosonic hopping amplitudes~\cite{KLMR11}, the next very interesting question might be 
whether the symmetry-protected topological (SPT) Haldane state~\cite{GW09,PTBO10}, observed in the extended BHM
(EBHM) with an additional nearest-neighbor particle repulsion~\cite{DBA06,BDGA08}, 
also shows up  in the extended AHM (EAHM).  
Because of its SPT order, the Haldane phase in the EBHM is separated 
from the topologically trivial MI phase by a phase transition, 
as long as the protecting symmetry---being a combination of bond-centered
inversion and a local unitary transformation---keeps up~\cite{PTBO10}. 
By breaking this symmetry, the two phases can be adiabatically connected 
without crossing a phase transition. Therefore, a sharp distinction 
between the two phases is only possible in the presence of 
the protecting symmetry, even though no spontaneous symmetry breaking occurs. 
As the hopping phase factor breaks the reflection parity 
in the system~\cite{Wilczek1990}, naively one might expect the
Haldane state to disappear in the EAHM for any finite fractional phase
$\theta$. However, this will not happen if the protecting symmetry
is appropriately generalized for finite $\theta$.

To comment on an anyonic topological Haldane state in one dimension,  we scrutinize  its protecting symmetry  
in the framework of the EAHM by analyzing the invariance of the density-dependent hopping
amplitudes (as for the EBHM in the limit $\theta\to0$).  
Calculating the generalized transfer matrices~\cite{PT12}
from the infinite matrix-product state (iMPS) of the infinite
density-matrix renormalization-group (iDMRG)~\cite{White92,Mc08,Sch11}
simulations, we prove the existence of the Haldane insulator (HI) state 
and derive the complete ground-state phase diagram of this paradigmatic anyonic model Hamiltonian at unit filling. 
In order to discriminate the topological HI phase from the other, more conventional 
Mott and density-wave (DW)  insulating phases in possible future experiments, 
we also determine the dynamical density response of the system, showing
a characteristic asymmetry in the Brillouin zone, which can be attributed 
to the fractional phase factor of the anyons.

The Hamiltonian of the 1D EAHM consists of three terms, 
$\hat{H}_{\rm EAHM}^{(a)}\equiv\hat{H}_t+\hat{H}_U+\hat{H}_V$, 
with 
\begin{eqnarray}
 \hat{H}_t=-t\sum_j
  \left(
   \hat{a}_j^{\dagger}\hat{a}_{j+1}^{\phantom{\dagger}}
   +{\rm H.c.}
  \right)\,,
\end{eqnarray} 
$\hat{H}_U=U\sum_{j}\hat{n}_j\left(\hat{n}_j-1\right)/2$ and
$\hat{H}_V=V\sum_{j}\hat{n}_j\hat{n}_{j+1}$,
describing the nearest-neighbor anyon transfer ($\propto t$), 
as well as the repulsive on-site ($\propto U$) and nearest-neighbor  ($\propto V$)
particle interaction, respectively. The anyon creation ($\hat{a}_{j}^{\dagger}$), annihilation ($\hat{a}_{j}^{\phantom{\dagger}}$) and particle 
number  ($\hat{n}_j=\hat{a}_j^\dagger\hat{a}_j^{\phantom{\dagger}}$) operators at lattice site $j$  are defined
by the generalized
commutation relations~\cite{Kundu99,KLMR11}:
\begin{eqnarray}
 \hat{a}_{j}^{\phantom{\dagger}}\hat{a}_{\ell}^{\dagger}
  -e^{-\mathrm{i}\theta\mathrm{sgn}(j-\ell)}
  \hat{a}_{\ell}^{\dagger}\hat{a}_{j}^{\phantom{\dagger}}
  &=&\delta_{j\ell}\, ,
  \\
 \hat{a}_{j}\hat{a}_{\ell}
  -e^{\mathrm{i}\theta\mathrm{sgn}(j-\ell)}a_{\ell}a_{j}
  &=& 0\, ,
 \label{CCR}
\end{eqnarray}
where the sign function $\mathrm{sgn}(j-\ell)=0$ for $j=\ell$ is mandatory,
since two anyons on the same site behave as ordinary bosons.
Anyons with $\theta=\pi$ represent so-called ``pseudofermions'', namely, 
they are fermions off-site, while being bosons on-site.

Performing a fractional Jordan--Wigner transformation~\cite{KLMR11},
\begin{equation}
\hat{a}_j=\hat{b}_j e^{\mathrm{i}\theta\sum_{\ell=1}^{j-1}\hat{n}_\ell}\,,
\label{ab-mapping}
\end{equation}
where $\hat{b}_j^\dagger$ ($\hat{b}_j^{\phantom{\dagger}}$)
is a boson creation (annihilation) operator,  
$\hat{H}_{\rm EAHM}^{(a)}$ becomes $\hat{H}_{\rm EAHM}^{(b)}$
with density-dependent hopping amplitudes,
\begin{eqnarray}
 \hat{H}_t=-t\sum_j
  \left(
   \hat{b}_j^{\dagger}\hat{b}_{j+1}^{\phantom{\dagger}}
   e^{\mathrm{i}\theta\hat{n}_j}
   +e^{-\mathrm{i}\theta\hat{n}_j}
   \hat{b}_{j+1}^{\dagger}\hat{b}_{j}^{\phantom{\dagger}}
  \right)\, .
  \label{Ht}
\end{eqnarray} 
That is when a boson hops to the left from site $j+1$ to site $j$
it acquires an occupation dependent phase $e^{\mathrm{i}\theta\hat{n}_j}$.
Of course, $\hat{n}_j=\hat{a}_j^\dagger\hat{a}_j^{\phantom{\dagger}}=\hat{b}_j^\dagger\hat{b}_j^{\phantom{\dagger}}$
which means that  $\hat{H}_U$ and $\hat{H}_V$ are form invariant
under the anyon-boson mapping~\eqref{ab-mapping}.

If we limit the maximum number of particles per site as $n_p=2$, the  EBHM, resulting in the limit 
$\theta\to 0$ from $\hat{H}_{\rm EAHM}^{(b)}$, maps to  an effective  $XXZ$
spin-1 chain~\cite{BDGA08}:
\begin{eqnarray}
 \hat{H}_{\rm eff}&=&
  -t\sum_j\left(\hat{S}_j^{+}\hat{S}_{j+1}^{-}+{\rm H.c.}\right)
  +\frac{U}{2}\sum_j\left(\hat{S}_j^z\right)^2 
  \nonumber \\
  &&+V\sum_j \hat{S}_j^z\hat{S}_{j+1}^z\, 
 \label{effectiveH}
\end{eqnarray}
with the pseudospin operator $\hat{S}_j^z=\hat{n}_j-1$.
Here we have neglected terms that break the particle-hole
symmetry. We note the negative sign of the first term compared 
to the usual $XXZ$ spin-chain Hamiltonian. 
This leads to a protecting modified inversion symmetry ${\cal I}^\prime$  
for the Haldane state of the EBHM~\cite{PTBO10}:
\begin{eqnarray}
{\cal I}^\prime=e^{\mathrm{i}\pi\sum_j\hat{S}_j^z}{\cal I}
 =e^{\mathrm{i}\pi\sum_j\left(\hat{n}_j-1\right)}{\cal I}\,. 
 \label{inversion-sym}
\end{eqnarray}
Owing to  the occupation-dependent hopping 
in \eqref{Ht} the HI phase in the EAHM seems not be protected by the modified inversion symmetry 
${\cal I}^\prime$.

To clarify whether $\hat{H}_t$ is invariant under certain symmetry operations, 
let us first consider the inversion symmetry operator  ${\cal I}$, acting on 
$\hat{H}_t \to \hat{H}_t^{\prime}
  ={\cal I}^{\phantom{\dagger}} \hat{H}_t {\cal I}^\dagger$ with
\begin{eqnarray}
 \hat{H}_t^{\prime}
  =-t \sum_j
  \left(
   \hat{b}_{j+1}^{\dagger}\hat{b}_{j}^{\phantom{\dagger}}
   e^{\mathrm{i}\theta\hat{n}_{j+1}}
   +e^{-\mathrm{i}\theta\hat{n}_{j+1}}
   \hat{b}_{j}^{\dagger}\hat{b}_{j+1}^{\phantom{\dagger}}
  \right)\, .
\end{eqnarray}
Applying next a time-reversal transformation ${\cal T}$, 
$\hat{H}_t^\prime \to \hat{H}_t^{\prime\prime} ={\cal T}\hat{H}_t^\prime {\cal T}^{-1}$,
we obtain 
\begin{eqnarray}
 \hat{H}_t^{\prime\prime}=-t \sum_j
  \left(
   \hat{b}_{j+1}^{\dagger}\hat{b}_{j}^{\phantom{\dagger}}
   e^{-\mathrm{i}\theta\hat{n}_{j+1}}
   +e^{\mathrm{i}\theta\hat{n}_{j+1}}
   \hat{b}_{j}^{\dagger}\hat{b}_{j+1}^{\phantom{\dagger}}
  \right)\, .
 \label{Hpp}
\end{eqnarray}
To see that $\hat{H}_t$ stays invariant under the combined symmetry operations,
we make the following transformation:
\begin{eqnarray}
  \hat{b}_j^{\dagger}
  &\to&
  e^{\mathrm{i}\theta\hat{n}_j(\hat{n}_j-1)/2}\, 
  \hat{b}_j^{\dagger}
  e^{-\mathrm{i}\theta\hat{n}_j(\hat{n}_j-1)/2}\, 
  =\hat{b}_j^{\dagger}e^{\mathrm{i}\theta\hat{n}_j}\, ,
  \label{T1}
  \\
 \hat{b}_j
  &\to&
  e^{\mathrm{i}\theta\hat{n}_j(\hat{n}_j-1)/2}\, 
  \hat{b}_j
  e^{-\mathrm{i}\theta\hat{n}_j(\hat{n}_j-1)/2}\, 
  =e^{-\mathrm{i}\theta\hat{n}_j}\hat{b}_j^{\phantom{\dagger}}\, .
  \label{T2}
\end{eqnarray}
Since the second term of Eq.~\eqref{Hpp} transforms as
$e^{\mathrm{i}\theta\hat{n}_{j+1}}
 \hat{b}_{j}^{\dagger}\hat{b}_{j+1}^{\phantom{\dagger}}
 \to
 \hat{b}_{j}^{\dagger}\hat{b}_{j+1}^{\phantom{\dagger}}
e^{\mathrm{i}\theta\hat{n}_{j}}$, it is equal to the first term of $\hat{H}_t$. 
\begin{figure}[tb]
 \begin{center}
  \includegraphics[clip,width=0.9\columnwidth]{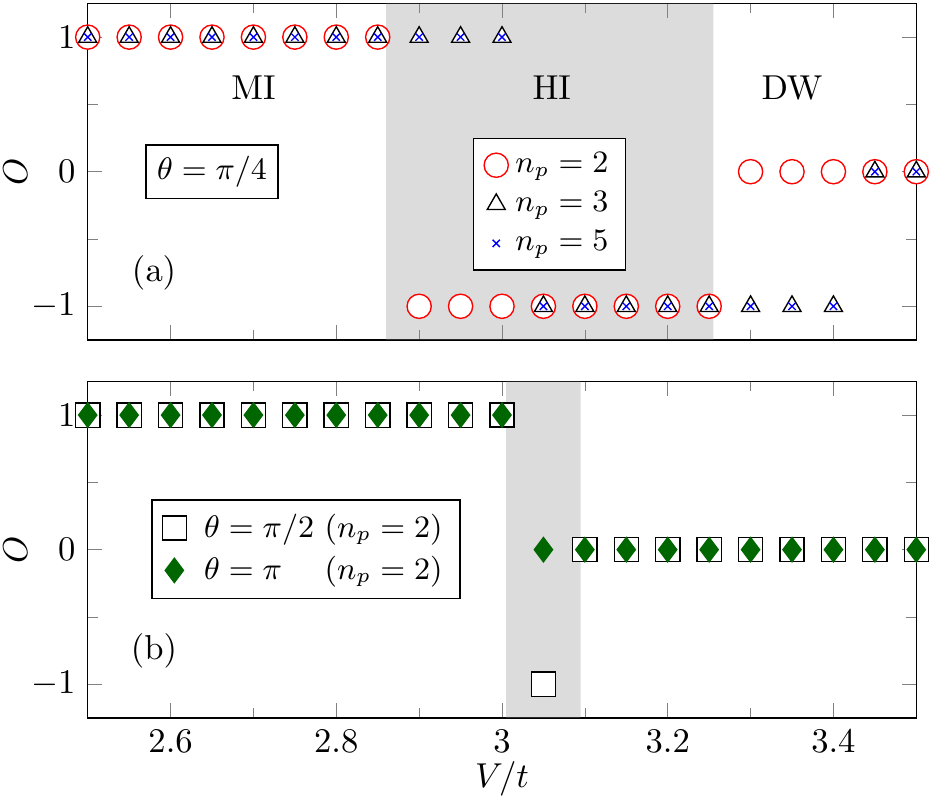}
 \end{center}
 \caption{(Color online). 
 Order parameter $O$, defined by Eq.~\eqref{topo-order-eq},  
 selecting  the topological state in the EAHM at fixed $U/t=5$
 and $\theta=\pi/4$ for different $n_p$ [panel (a)],
 and at fixed $\theta=\pi/2$ and $\pi$ for $n_p=2$
 [panel (b)]. Data obtained by iDMRG calculations
 with a (relatively small) bond dimension $\chi=100$.}
 \label{topo-order}
\end{figure}
Therefore the Hamiltonian $\hat{H}_{\rm EAHM}^{(b)}$ is invariant under the transformation
\begin{eqnarray}
{\cal K}
 =e^{\mathrm{i}\theta\sum_j\hat{n}_j(\hat{n}_j-1)/2}
{\cal I}\,{\cal T}\, .  
\end{eqnarray}

We now show that the combination of 
${\cal R}^z=e^{\mathrm{i}\pi\sum_j\hat{S}_j^z} = e^{\mathrm{i}\pi\sum_j\left(\hat{n}_j-1\right)}$ 
and ${\cal K}$ is related to an SPT phase 
in the EAHM, and define a corresponding topological order parameter. 
Following Ref.~\cite{Vidal07}, we use the iMPS
representation formed by
complex $\chi\times\chi$ matrices $\Gamma_\sigma$ and a positive, real, diagonal matrix $\Lambda$: 
\begin{eqnarray}
|\psi \rangle = \ \sum_{\mathclap{...\sigma_j,\sigma_{j+1}...}} \ ...\;\Lambda \Gamma_{\sigma_j} \Lambda \Gamma_{\sigma_{j+1}} ... \; |...,\sigma_j,\sigma_{j+1},...\rangle \,,
\end{eqnarray}
where the index $\sigma$ labels the basis states of the local Hilbert spaces. 
The iMPS is assumed to be in the canonical form: 
$\sum_\sigma\Gamma_\sigma^{\phantom{\dagger}}\Lambda^2\Gamma_\sigma^{\dagger} 
 = \sum_\sigma\Gamma_\sigma^{\dagger}\Lambda^2\Gamma_\sigma^{\phantom{\dagger}} =
 \mathbbm{1}$. 
If a state $|\psi\rangle$ is invariant under an internal symmetry that is represented by a unitary matrix $\Sigma_{\sigma\sigma^\prime}$, then the transformed $\Gamma_\sigma$ matrices satisfy~\cite{PWSVC08,PTBO10}
\begin{eqnarray}
  \sum_{\sigma^\prime}\Sigma_{\sigma\sigma^\prime}\Gamma_{\sigma^\prime}
   =e^{\mathrm{i}\varphi}U^{\dagger}
   \Gamma_\sigma^{\phantom{\dagger}} U^{\phantom{\dagger}}\,,
   \label{MPS-symmetry}
\end{eqnarray}
where $U$ is a unitary matrix that commutes with $\Lambda$ 
 and $e^{\mathrm{i}\varphi}$ is a phase factor. 
Similar relations hold for time reversal symmetry, inversion symmetry, and a combination of both. In those cases $\Gamma_\sigma$ on the left-hand side is replaced by its complex conjugate $\Gamma_\sigma^*$, its transpose $\Gamma_\sigma^T$ and its Hermitian transpose $\Gamma_\sigma^\dagger$, respectively. 
The properties of the matrices $U$ can be used to classify SPT phases~\cite{PTBO10,CGW11}. 
For instance, in the case of time reversal or (modified) inversion symmetry the matrices satisfy $U_{\cal T}^{\phantom{*}}U_{\cal T}^* = \pm \mathbbm{1}$ or $U_{{\cal I}^{(\prime)}}^{\phantom{*}}U_{{\cal I}^{(\prime)}}^{*} = \pm \mathbbm{1}$, and the sign distinguishes between two symmetric phases. 
In the EAHM, the situation is slightly different because time reversal
and inversion are not symmetries of the system, only a combination ${\cal K}$ of them is. 
For ${\cal R}^z$ and ${\cal K}$ we have 
$U_{{\cal R}^z}^2=e^{\mathrm{i}\alpha_{{\cal R}^z}} \mathbbm{1}$ 
and $U_{{\cal K}}^2=e^{\mathrm{i}\alpha_{{\cal K}}} \mathbbm{1}$. 
From this we can derive an SPT order parameter similar to the case of the $Z_2 \times Z_2$ spin rotation symmetry of ${\cal R}^z$ and ${\cal R}^x$ in the spin-1 $XXZ$ chain~\cite{PTBO10}. 
Since the phase factors 
$e^{\mathrm{i}\alpha_{{\cal R}^z}}$ 
and 
$e^{\mathrm{i}\alpha_{{\cal K}}}$ 
can be removed by absorbing them into the corresponding matrices $U_{{\cal R}^z}$ and $U_{{\cal K}}$ they have no physical meaning. However, if both ${\cal R}^z$ and ${\cal K}$ are preserved, the combination ${\cal R}^z{\cal K}$ is a symmetry as well and its phase factor is not arbitrary if $U_{{\cal R}^z}$ and $U_{{\cal K}}$ have been fixed. Indeed one can show that $U_{{\cal R}^z}U_{{\cal K}} = \pm U_{{\cal K}}U_{{\cal R}^z}$ which defines two different phases. 
To verify that the EAHM has a nontrivial topological phase protected by ${\cal R}^z$ and ${\cal K}$, we calculate the order parameter~\cite{PT12}
\begin{eqnarray}
 O=\frac{1}{\chi}\mbox{tr}
   \left(
    U_{{\cal K}}^{\phantom{\dag}}
    U_{{\cal R}^z}^{\phantom{\dag}}
    U_{{\cal K}}^{\dag}
    U_{{\cal R}^z}^{\dag}
   \right)\, ,
   \label{topo-order-eq}
\end{eqnarray}
if the state is symmetric under both ${\cal K}$ and ${{\cal R}^z}$. 
Otherwise, if one of the symmetries is broken, the order parameter is zero. 

The iDMRG results for the order parameter are shown in Fig.~\ref{topo-order}. 
If $U_{{\cal K}}$ and $U_{{\cal R}^z}$ commute ($O=1$),
the system is in a trivial phase, i.e., a site-factorizable MI state, whereas if they anticommute ($O=-1$), 
the system realizes a nontrivial HI state. Since the order parameter $O$ changes its sign only if a phase transition
takes place, the HI is a well-defined phase of the EAHM. 
Increasing the number of particles per site $n_p$ at fixed $U/t=5$, the HI phase ($O=-1$) slightly 
shifts to larger value of $V/t$ but, most notably,  the Haldane phase still occupies 
a solid parameter region, see the data for $n_p=3$ and $5$
in Fig.~\ref{topo-order}(a).
Increasing the fractional angle $\theta$ for $n_p=2$, the Haldane state
region narrows [see Fig.~\ref{topo-order}(b) for $\theta=\pi/2$]
and disappears  (at least) for $\theta=\pi$~\cite{Comment}. 
We would like to emphasize that the  HI sector marked in Fig.~\ref{topo-order} by the gray area 
agrees with that extracted from the correlation length, the entanglement
spectrum, and the numerically 
obtained central charge~\cite{SuppMat}.

\begin{figure}[t]
 \begin{center}
  \includegraphics[clip,width=0.9\columnwidth]{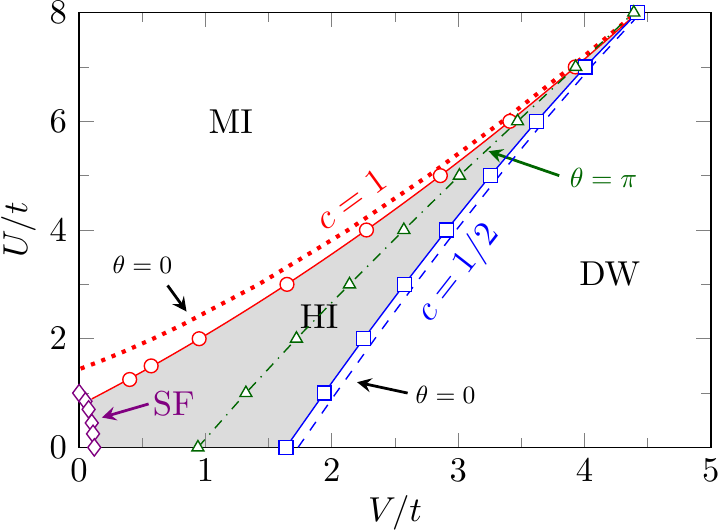}
 \end{center}
 \caption{(Color online). Ground-state phase diagram of the 
 extended anyon-Hubbard model in one dimension, where the particle density    
 $\rho=1$, $n_p=2$, and $\theta=\pi/4$. Most notably the Haldane insulator (HI), 
 located between Mott insulator (MI) and density wave (DW) insulating
 phases in the EBHM, survives for any $\theta>0$, i.e., in the anyonic
 case. Likewise the superfluid (SF) appears in the very weak-coupling regime. 
 The MI-HI (squares) and HI-DW (circles) transition points 
 can be determined by a divergent correlation length $\xi_\chi$
 as $\chi$ increases, i.e., the model becomes critical with
 the central charge $c=1$ and $c=1/2$, 
 respectively (see Ref.~\cite{SuppMat}).
 For comparison, the dotted (dashed) line marks the MI-HI (HI-DW) transition
 in the EBHM ($\theta=0$)~\cite{ELF14}. The dash-dotted line with triangles up denotes the
 first-order MI-DW phase transition for $\theta=\pi$.
 }
\label{pd}
\end{figure}

\begin{figure*}[t!]
 \begin{center}
  \includegraphics[clip,width=1.9\columnwidth]{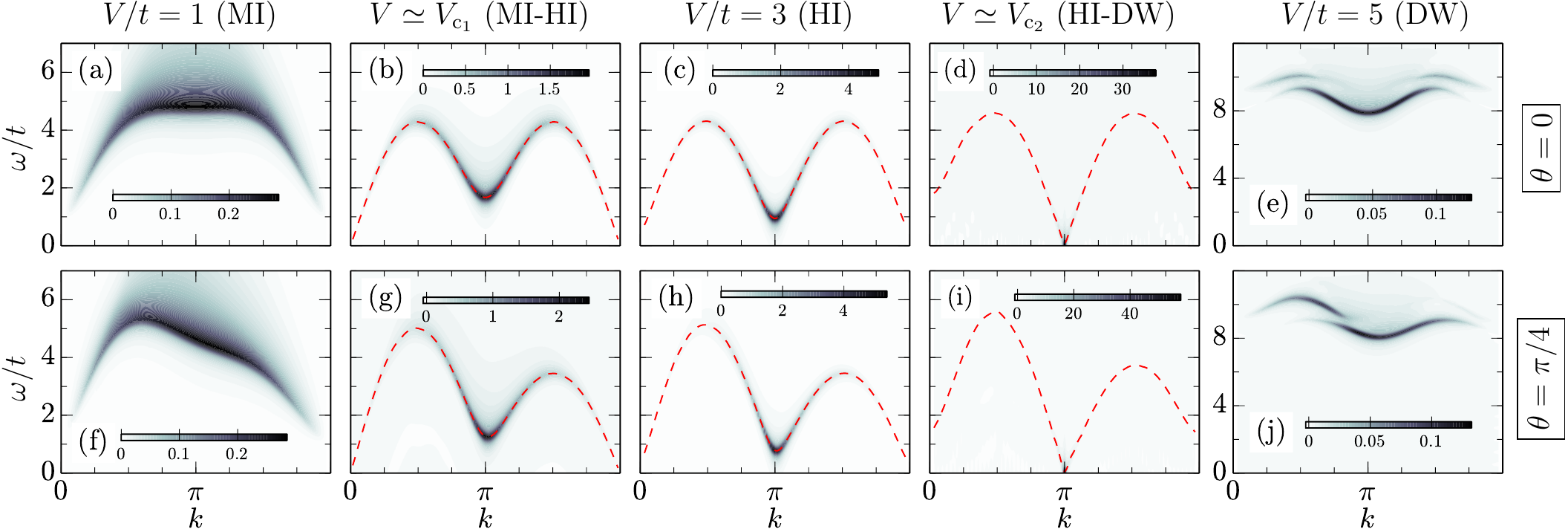}
 \end{center}
 \caption{(Color online). Intensity plots of the dynamical structure
 factor $S(k,\omega)$ in the EBHM ($\theta=0$, upper panels)
 and in the EAHM ($\theta=\pi/4$, lower panels)
 for characteristic values of $V/t$ at fixed $U/t=5$. Again the maximum number of particles per
 site is limited to $n_p=2$. Dashed lines in panels (b)-(d) and (g)-(i)
 mark the highest intensity of $S(k,\omega)$ in the $k$-$\omega$ plane.
 }
\label{Skw}
\end{figure*}

Figure~\ref{pd} represents the ground-state phase diagram of the 1D EAHM in the $V$-$U$ plane, 
as obtained from large-scale iDMRG calculations for $\theta=\pi/4$ and $n_p=2$. The phase 
boundaries are determined simulating the order parameter $O$, as well as the 
correlation length and the entanglement spectrum~\cite{SuppMat}.  
The EAHM exhibits  three different insulating phases (MI, DW, and HI) and a superfluid state in the weak 
interaction regime, just as for the EBHM~\cite{ELF14} but with the addition that the region of the intervening anyonic HI phase at $\theta=\pi/4$ is slightly reduced. 
The HI vanishes in the pseudofermionic case  ($\theta=\pi$). 
According to field theory for the EBHM~\cite{LLY07,BDGA08}, which is based on the 
bosonization procedure developed for integer-spin chains~\cite{Schulz86,Tsvelik90}, the MI-HI and HI-DW quantum phase transitions belong  
to the universality class of Tomonaga-Luttinger liquid and Ising model, with central charge $c=1$ and $1/2$, respectively,
see Fig.~S1(c) in~\cite{SuppMat}. That is  the universality classes are not  modified by the fractional angle.

Perhaps the most  striking feature of the AHM is the asymmetry of the
momentum distribution function in $k$-space~\cite{HZC09,TEP15}. 
The position of the maximum strongly depends on the fractional phase $\theta$ 
[remind that the momentum distribution diverges  at $k=0$ in the BHM ($\theta=0$)]. 
We expect  that this asymmetry can also be observed in dynamical quantities such as the dynamical structure
factor $S(k,\omega)$. Hence, if an anyonic system will be realized in optical lattices,  $S(k,\omega)$ might be
one of the best physical quantities to look at, comparing theoretical predictions with real experiments, like for 1D Bose--Hubbard type models~\cite{CFFFI09}. $S(k,\omega)$ should be easily accessible by momentum resolved Bragg spectroscopy~\cite{EGKPLPS10}.  Furthermore,  it has been recently demonstrated that $S(k,\omega)$ can also be used to distinguish the topological HI from the conventional MI and DW states~\cite{ELF14,EF15b}, 
in analogy to exploiting the dynamical spin-spin structure
factor in the spin-1 $XXZ$ chain~\cite{EF15}.

The dynamical density structure factor is defined as
\begin{eqnarray}
 S(k,\omega)=
  \sum_n \left|\langle \psi_n|\hat{n}_k|\psi_0\rangle\right|^2
  \delta(\omega-\omega_n)\, , 
  \label{Skw-eq}
\end{eqnarray}
where $|\psi_0\rangle$ ($|\psi_n\rangle$) denotes the ground 
($n$th excited) state, and $\omega_n=E_n-E_0$. 
To compute this quantity, we follow Ref.~\cite{PVM12} and first determine the two-point correlation 
function $\langle\psi_0|\hat{n}_j(\tau)\hat{n}_0(0)|\psi_0\rangle$
by real-time evolution of the iMPS $|\psi_0\rangle$. Fourier transformation 
then provides us with accurate numerical results of the dynamical structure 
factor in the EAHM. 

Figure~\ref{Skw} compares the intensity of the dynamical 
wave-vector-resolved density response in the EBHM ($\theta=0$) 
with those in the EAHM for $\theta=\pi/4$, for $U/t=5$, at five characteristic $V/t$-values. 
One point worthy of remark is that each of the phases and phase transitions 
can be distinguished by looking at $S(k,\omega)$. 
In the MI, at $V=t$ [Figs.~\ref{Skw}(a) and \ref{Skw}(f)],
the excitation gap appears 
at $k\approx 0$. With increasing $V/t$, the MI-HI transition occurs
at $V\simeq V_{\rm c_1}$, where the excitation gap closes at $k=0$,
as shown in Figs.~\ref{Skw}(b) and \ref{Skw}(g). Deep in the HI phase, $V=3t$ 
[Figs.~\ref{Skw}(c) and \ref{Skw}(h)],
the spectral weight exclusively concentrates 
at $k\simeq\pi$, and there are finite excitation gaps at $k=0$ and $\pi$. 
It is of particular interest to see whether the gap $S(k,\omega)$
closes at the HI-DW transition point. 
Indeed, the excitation gap
at $V=V_{\rm c_2}$ closes, but at momentum $k=\pi$, reflecting 
the lattice-period doubling in the DW phase. Moreover, in the DW phase 
[Figs.~\ref{Skw}(e) and \ref{Skw}(j)], we find a large excitation gap 
at $k=\pi$ and two dispersive branches, where a changeover 
of the intensity maximum occurs at $k=\pi/2$ ($k=3\pi/4$)
for $\theta=0$ ($\theta=\pi/4$).  
Interestingly, the influence of the occupation-dependent 
phase of $\hat{H}_t$ in Eq.~\eqref{Ht} shows up in $S(k,\omega)$ as well, 
which helps to differentiate the results  from those of the EBHM. 
$S(k,\omega)$ of the EAHM is asymmetric for any $0<\theta<\pi$, 
while $S(k,\omega)$ in the EBHM is always symmetric about $k=\pi$.

To summarize, we carried out an unbiased numerical investigation of the extended anyon-Hubbard model in one dimension
and determined its ground-state phase diagram with high precision exploiting the behavior of correlation lengths and entanglement spectra. Defining an order parameter that distinguishes trivial and nontrivial topological phases, we were able to show that the
EAHM  possesses an anyonic Haldane insulator state sandwiched between superfluid, Mott insulator and density-wave phases. 
Both the HI-MI and HI-DW quantum phase transitions are critical with central charge 1 and 1/2, respectively.  While the HI state 
survives the EBHM limit $(\theta=0$), it vanishes when the system is
composed of pseudofermions  $(\theta=\pi$). If a 1D interacting anyonic
system could be realized experimentally in the future, maybe in an
optical-lattice setup with ultracold atoms,  we suggest performing
momentum-resolved Bragg spectroscopy to look for the pronounced
asymmetry of the density response spectra in momentum space
that we have demonstrated in our model calculation theoretically.

{\it Acknowledgments}.---The iDMRG simulations were performed
using the ITensor library~\cite{ITensor}. This work was supported
by Deutsche Forschungsgemeinschaft (Germany), SFB 652, project B5.


\begin{thebibliography}{33}
\bibitem [{Com()}]{Comment}%
  \BibitemOpen
  \href@noop {} {}\bibinfo {note} {Note that our iDMRG scheme holds for
  arbitrary $\theta$. Additional iDMRG calculations confirm that the HI phase
  exists not only for the specific fractional angles $\theta$ used  in
  Fig.~\ref{topo-order} but also for other values of $\theta \in (0,\pi)$ (not shown). The 
  critical $\theta_c$, where the HI phase disappears, depends on the model parameters 
  $U$, $V$, and $n_p$.}\BibitemShut {Stop}%
\end{thebibliography}
\bibliographystyle{apsrev4-1}

\clearpage
\appendix 
\section{\Large 
  Supplementary material}
\renewcommand{\theequation}{$\text{S}$\arabic{equation}}
\setcounter{equation}{0}
\renewcommand{\thefigure}{$\text{S}$\arabic{figure}}
\setcounter{figure}{0}

\setcounter{page}{1}
\makeatletter
\renewcommand{\bibnumfmt}[1]{[S#1]}
\renewcommand{\citenumfont}[1]{S#1}

As discussed in the main text, we find compelling evidence for the existence 
of the symmetry-protected topological  Haldane insulator phase 
in the one-dimensional (1D) extended anyon-Hubbard model (EAHM), 
by calculating an order parameter from the largest eigenvalues of the generalized transfer matrix
within an infinite density-matrix renormalization group (iDMRG) scheme.

Here we show that for the  EAHM (with maximum number of particles per site $n_p=2$) 
further quantities can be exploited in order to determine and characterize the phase boundaries and 
quantum phase transitions with high precision.

 The entanglement analysis in particular provides us with valuable information 
 about the existence of a symmetry protected Haldane insulator (HI) in the EAHM. 
 Furthermore, it allows to determine the phase boundaries between the HI and other insulating phases.
Dividing a system into two subblocks, 
$\cal{H}=\cal{H}_{\textrm L}\otimes\cal{H}_{\textrm R}$,
and considering the reduced density matrix 
$\rho_{\textrm L}=\mathrm{Tr}_{\textrm R}[\rho]$,
the entanglement spectra~\cite{LH08supp} can be extracted from 
the singular values $\lambda_\alpha$ of $\rho_{\textrm L}$ as
$\epsilon_\alpha=-2\ln\lambda_\alpha$.
In addition, the correlation length $\xi_\chi$ can be obtained from
the second largest eigenvalue of the transfer matrix for some
fixed bond dimension $\chi$ in an iDMRG simulation~\cite{Mc08supp,Sch11supp}.
While $\xi_\chi$ stays finite as a consequence of the fixed bond dimension
$\chi$,  the physical correlation length will diverge at the
critical point. Nevertheless, $\xi_\chi$ is useful to pinpoint a phase boundary  
because it rapidly increases with $\chi$ close to
the quantum phase transition point,
see also Ref.~\cite{VMC04} for the corresponding discussion 
in the AKLT model~\cite{AKLT87}.

\begin{figure}[t!]
 \begin{center}
  \includegraphics[clip,width=0.95\columnwidth]{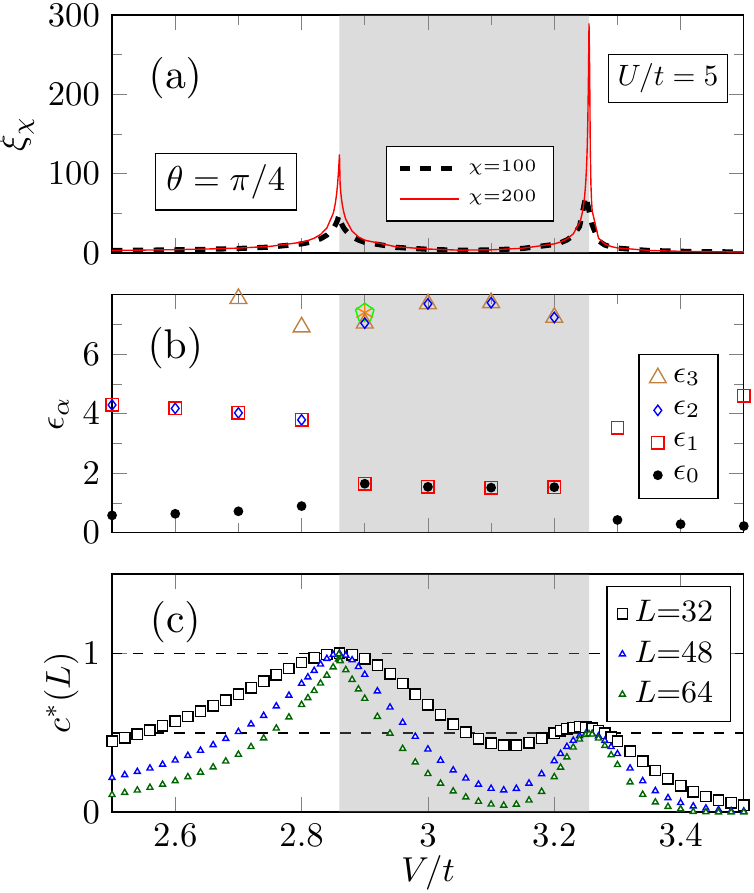}
 \end{center}
 \caption{(Color online). 
 Correlation length $\xi_\chi$ [panel (a)] and entanglement spectrum
 $\epsilon_\alpha$ [panel (b)] as a function of $V/t$ for $U/t=5$
 and $\theta=\pi/4$ from iDMRG. 
 Panel (c) displays the central charge $c^\ast(L)$ for the same
 parameter set, signaling a MI-HI (HI-DW) 
 quantum phase transition with $c=1$ ($c=1/2$). 
 Here data obtained by the finite-system DMRG with periodic boundary conditions.
 }
\label{xi-es-cstar}
\end{figure}

Figures~\ref{xi-es-cstar}(a) and (b) show $\xi_\chi$ 
and $\epsilon_\alpha$ as functions of $V/t$ for fixed $U/t=5$. 
The strong upturn of $\xi_\chi$ indicates the formation of a HI phase
in the EAHM for $\theta>0$. We find distinct peaks at $V_{{\rm c}_1}\simeq2.859t$ and
$V_{{\rm c}_2}\simeq3.255t$, which become more pronounced as $\chi$ grows
from 100 to 200, signaling a divergence of $\xi_\chi\to\infty$
as $\chi\to\infty$. At the same time, the entanglement spectra develops a characteristic double
degeneracy in all entanglement levels for $V_{{\rm c}_1}< V < V_{{\rm c}_2}$, indicating a 
symmetry-protected topological phase between MI and DW states.

The universality class of these quantum phase transitions can be 
explored by calculating the central charge numerically,  
just as in case of the EBHM~\cite{ELF14supp}. When the system gets 
critical the central charge can be determined very accurately by DMRG,
utilizing  the relation~\cite{Ni11}
\begin{eqnarray}
 c^\ast(L) \equiv \frac{3[S_L(L/2-1)-S_L(L/2)]}{\ln[\cos(\pi/L)]}\;.
 \label{cstar}
\end{eqnarray}
In this way, the MI-SF transition in the BHM~\cite{EFGMKAL12}, and especially,
the university class of the MI-HI and HI-DW quantum phase transitions
in the EBHM have been determined in the past~\cite{ELF14supp}.

Figure~\ref{xi-es-cstar}(c) displays $c^\ast(L)$ for the 1D EAHM, where the model
parameters are the same as in Figs.~\ref{xi-es-cstar}(a) and (b). 
Running the  DMRG we adopt periodic boundary conditions for system sizes up to $L=64$.
For $U/t=5$ and $V\simeq V_{{\rm c}_1}$ [$V\simeq V_{{\rm c}_2}$],
we find $c^\ast(L=64)\simeq0.996$ [$c^\ast(L=64)\simeq0.494$],
which points to the universality class of the Luttinger liquid
(Ising) model, in accordance with what was obtained for the corresponding quantum phase transitions
in the EBHM ($\theta=0$).

\end{document}